\documentclass{LT23auth}
\usepackage{graphicx}
\begin{document}
\begin{frontmatter}
%
\title{Superconducting SET  
with tunable electromagnetic environment}
\author{Michio Watanabe\thanksref{thank1}},
\author{Koji Ishibashi}, 
\author{Yoshinobu Aoyagi}
\address{Semiconductors Laboratory, RIKEN, and CREST-JST, 
2-1 Hirosawa, Wako-shi, Saitama 351-0198, Japan}
%
\vspace*{-1.5\baselineskip}   
%
\thanks[thank1]{Corresponding author.  
 E-mail: michio@postman.riken.go.jp}
%
\begin{abstract}
We have studied the environmental effect on  
superconducting single-electron transistors (S-SETs)  
by biasing S-SETs with 
arrays of small-capacitance dc SQUIDs, 
whose effective impedance can be varied {\itshape in situ}.   
As the zero-bias resistance of the arrays is increased, 
Coulomb blockade in the S-SET becomes sharper, 
and the gate-voltage dependence 
changes from 
$e$-periodic to 2$e$-periodic.  
The SQUID arrays could be used as on-chip noise filters.  
\end{abstract}
%
%
%
\end{frontmatter}
%
\vspace*{-3.5\baselineskip}   
%
\section{Introduction}
%
\vspace*{-\baselineskip}   
%
Small-capacitance Josephson junction 
has been playing an important role in studying the 
interplay between quantum mechanically conjugate valuables, 
the Josephson phase and the charge on the island 
electrode~\cite{QPT}.  
It is also a promising candidate 
for solid-state realizations of quantum 
computing~\cite{qubit}.
In this system, coupling to an electromagnetic environment 
is a key issue.  The coupling influences 
the interplay between the phase and the charge.  
Moreover, it is a major source of decoherence 
in quantum bits.  

In this work we study the environmental effect on  
superconducting single-electron transistors (S-SETs), 
which are building blocks of nano-circuits.   
The effects of dissipation on S-SETs have been 
investigated by capacitively coupling a two-dimensional 
electron gas to an S-SET~\cite{SSET2DEG}.    
We take a different approach as described 
in the next section.  
%
\vspace*{-1.5\baselineskip}   
%
\section{Experiment}
\begin{figure}[btp]
\begin{center}\leavevmode
\includegraphics[width=0.8\linewidth,clip]{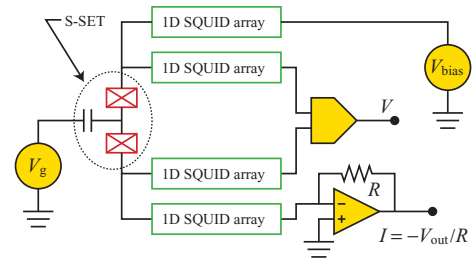}
\caption{ 
Schematic diagram of the sample and the circuit.
}\label{fig:diagram}\end{center}\end{figure}

%
\vspace*{-\baselineskip}   
%
We have biased 
S-SETs with one-dimensional (1D) arrays of 
small-capacitance dc superconducting quantum interference 
devices (SQUIDs) as shown in Fig.~\ref{fig:diagram}.   
%
%
With the SQUID configuration, the effective impedance 
of the arrays can be varied {\itshape in situ} 
by applying a weak ($1-10$~mT) external magnetic field $B$, 
i.e., the arrays are tunable 
environment~\cite{Wat01PRL,Wat01SUST,Wat02} for the SET.  
For a scanning electron micrograph of a SQUID array, 
see Fig.~1 of Ref.~\cite{Wat02}.  
We measured the SET in a four-point configuration, 
where the bias was applied through one pair of SQUID leads  
and the potential difference was probed through the other 
pair of leads.   
The SQUID arrays could be measured in a two-point 
configuration on the same side of the SET, 
and the zero-bias resistance $R'_0$ obtained in this 
measurement may characterize the environment~\cite{Wat01PRL}.  

The samples consist of Al/Al$_2$O$_3$/Al junctions 
fabricated on a SiO$_2$/Si substrate.  
The area of each junction in the SET 
is $0.1\times0.1$~$\mu$m$^2$, so that the charging energy 
of the SET is on the order of 1~K.  
%
\vspace*{-1.5\baselineskip}   
%
\section{Results and discussion}
\begin{figure}[btp]
\begin{center}\leavevmode
\includegraphics[width=0.875\linewidth,clip]{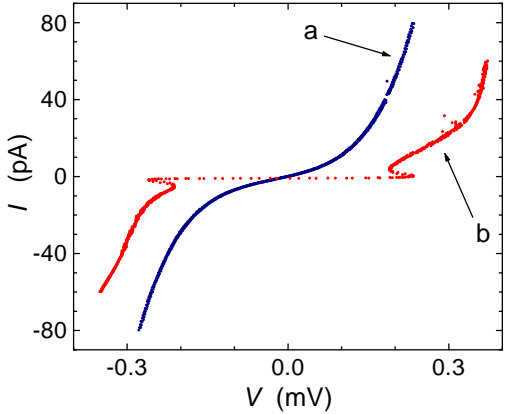}
\caption{ 
Current-voltage characteristics of a superconducting 
SET at $T=0.02$~K in different environments.  
}\label{fig:IV}\end{center}\end{figure}

%
\vspace*{-\baselineskip}   
%
We show in Fig.~\ref{fig:IV} low-temperature ($T=0.02$~K) 
current-voltage ($I$-$V$) curves of an S-SET 
in different environments, 
$R'_0=0.2$~M$\Omega$ ($B=0$) for a  
and $R'_0=0.3$~G$\Omega$ ($B=6.8$~mT) for b.  
%
%
Here, the temperature is 0.02~K 
and the gate voltage, $V_{\rm g}$, is 0.  
For the sample in Fig.~\ref{fig:IV}, 
the normal-state resistance 
of the SET is 0.10~M$\Omega$,   
so that the Josephson energy is 
estimated to be 0.1~K.   
The parameters of the SET should be the same 
for the two curves in Fig.~\ref{fig:IV}, 
because the SET does not have a SQUID 
configuration and the field applied here is much smaller 
than the critical field for Al films ($\approx 0.1$~T).  
The environment for the SET (the SQUID arrays), however, 
is strongly varied with $B$.  
The behavior of the SET demonstrated in Fig.~\ref{fig:IV} 
does not result from the magnetic-field influencing 
the $I$-$V$ curve of the SET, but rather from an environmental 
effect on the SET.  
While Coulomb blockade in curve~a is rounded,  
that in curve~b is sharp.  
Furthermore, curve~b has a `back-bending' in the low-current part, 
which is evidence of coherent single-Cooper-pair 
tunneling~\cite{Wat01SUST}.  
For the difference between curves~a and b 
quasiparticles generated by external noise 
may be responsible, because   
no filter has been installed in or at the top of our cryostat.  
A sufficiently large $R'_0$ 
would weaken the coupling to the noise source and suppresses 
the quasiparticle generation.  
 
\begin{figure}[btp]
\begin{center}\leavevmode
\includegraphics[width=0.95\linewidth,clip]{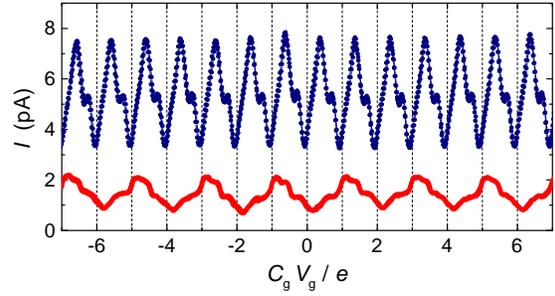}
\caption{ 
Current vs. normalized gate voltage at $T=0.02$~K 
for the same 
superconducting SET as in Fig.~\ref{fig:IV}.  
The upper (lower) data set was taken in the same 
environment as for curve~a (b) in Fig.~\ref{fig:IV}.  
}\label{fig:gate}\end{center}\end{figure}

We have also examined the dependence on $V_{\rm g}$ 
as shown in Fig.~\ref{fig:gate}.  
We fixed $V_{\rm bias}$ (see Fig.~\ref{fig:diagram}) 
and measured the current $I$ as a function of $V_{\rm g}$.  
When $V$, the the potential difference across the SET 
is large compared to the voltage drop at the other parts of the circuit, 
$V$ is nearly independent of $V_{\rm g}$.    
This is the case for the upper data set in Fig.~\ref{fig:gate} 
($R'_0=0.2$~M$\Omega$, $B=0$), and $V=0.1$~mV.  
In the opposite case, $V$ also oscillates as a function of 
$V_{\rm g}$, which makes the amplitude of the current oscillation 
small.  In the case of the lower data set in Fig.~\ref{fig:gate} 
($R'_0=0.3$~G$\Omega$, $B=6.8$~mT), $V$ oscillates between 
0.1 and 0.3~mV.   
Nevertheless, it is still possible to clearly observe 
the oscillation of $I$.    
In Fig.~\ref{fig:gate}, $V_{\rm g}$ is normalized by $e/C_{\rm g}$, 
where the gate capacitance $C_{\rm g}=6$~aF 
is estimated from a similar gate-sweep measurement in a strong enough 
magnetic field ($B=0.1$~T) to drive the sample into the normal state.  
By tuning the environment, the $V_{\rm g}$ dependence 
changes from $e$-periodic to $2e$-periodic, 
i.e., the dominant tunneling process changes from 
quasiparticle tunneling to Cooper-pair tunneling.  
This observation is consistent with the picture in the end of 
the preceding paragraph that the quasiparticle generation  
is suppressed by the SQUID arrays when the arrays 
have a sufficiently large $R'_0$.  
%
%
\vspace*{-1.5\baselineskip}   
%
\section{Conclusion}
%
\vspace*{-\baselineskip}   
%
We could sharpen Coulomb blockade in the S-SET    
and change the dependence on the gate voltage  
from $e$-periodic to 2$e$-periodic,   
by tuning the electromagnetic environment 
composed of small-capacitance SQUID arrays 
in the immediate vicinity of an S-SET.   
The SQUID arrays behaved as effective 
noise filters.
%
\vspace*{-1.5\baselineskip}   
%
\begin{ack}
%
\vspace*{-\baselineskip}   
%
This work was supported in part by 
Special Postdoctoral Researchers Program 
and President's Special Research Grant of RIKEN.  
\end{ack}
%
%

%
\end{document}